**Social Media Usage in Kuwait: A Comparison of Perspectives between Healthcare Practitioners and Patients**

Fatemah Husain

Kuwait University

f.husain@ku.edu.kw

Vivian Motti

George Mason University

vmotti@gmu.edu

# Social Media Usage in Kuwait: A Comparison of Perspectives Between Healthcare Practitioners and Patients

## Abstract

Social Media has been transforming numerous activities of everyday life, impacting also healthcare. However, few studies investigate the medical use of social media by patients and medical practitioners, especially in the Arabian Gulf region and Kuwait. To understand the behavior of patients and medical practitioners in social media toward healthcare and medical purposes, we conducted user studies. Through an online survey, we identified a decrease in patients' and medical practitioners' use of social media for medical purposes. Patients reported to be more aware than practitioners concerning: health education, health-related network support, and communication activities. While practitioners use social media mostly as a source of medical information, for clinician marketing and for professional development. The findings highlighted the need to design a social media platform that support healthcare online campaign, professional career identity, medical repository, and social privacy setting to increase users' engagements toward medical purposes.

## Introduction

Social media has become part of daily-life activities for human users regardless of their locations. Recent reports (Chaffey, 2018) estimate that 3.196 billion users accessed social media as of January 2018. With social media access rates being proportional to Internet access, the use is higher in Europe (90%) and the US (88%), and lower in middle Africa (12%) and Southern Asia (36%) (Kemp, 2018). Such a large adoption rate impacts users' lives across numerous activities, suggesting the potential of social media to significantly impact everyday life.

The integration of social media into healthcare sector can provide great opportunities in solving major problems in healthcare services. The current healthcare system in Kuwait does not support online direct communication between patients and healthcare providers, and the number of physicians and service providers is not increasing in proportion to the number of patients and population, suggesting that social media has potential to bridge the gap between patient needs and healthcare service providers in a timely basis, with benefits regardless of geographic location, professional background, and economic status. Especially concerning the medical appointments that do not require physical diagnosis, such as urgent inquiries and healthcare education, as those are the most suitable to be replaced with social media.

Positive applications of social media in serving healthcare delivery are broad on their effects, impacting organizations, professionals, and patients. In the US, 25% of healthcare organizations are adopting social media by creating accounts representing their organizations to maintain their online reputation among their competitors and customers (McCaughey et al., 2014), and as a convenient organizational communication tool to connect with the public (Edosomwan et al., 2011). The large adoption of diverse social media channels has direct positive impacts on the overall hospital rating and on the willingness to recommend the hospital (McCaughey et al., 2014). Medical practitioners can practice their professional role online, promote knowledge sharing (Chaudhry, 2013), promote professional development (Bernhardt et al., 2014), and educate themselves as well as others about health-related issues (Brokowski, 2009; Bernhardt et al., 2014; Ventola, 2014). The use of social media by medical practitioners enables them to update their knowledge and keep them connected with international peers. Social media empowers patients with information by connecting them with healthcare providers, learning about a condition, treatments available, and connecting them with other patients in similar situations to receive peer and emotional support (Taiminen, 2016). However, social media also has adverse effects on healthcare when misinformation is spread, and patient privacy is violated. To harness the benefits that social media applications can have on patients' lives and for healthcare practitioners as well, it is very important to understand current user behavior and patient concerns related to medical content and social media.

Detailed investigations into user behaviors have potential to create future solutions by emphasizing on medical purpose supportive features and reducing possible privacy risks.

This study focuses on understanding user behaviors toward medical and healthcare usage of social media in Kuwait. The main research questions of this study are:

(1) What are the profile of users who adopt social media for healthcare (including their demographic information, current usage of social media, privacy settings and potential concerns)?

(2) How do users employ social media applications for medical purposes (concerning the channels used for medical purposes, type of the medical contents used and benefits of using social media)?

This paper has four main sections. The first section provides background information on the healthcare system in Kuwait. The second section discusses the benefits and challenges of the use of social media by the patients and medical practitioners. The third section presents and justifies the methodology used for collecting and analyzing the data. The fourth section presents the results of the data analysis. The paper closes with a discussion of the results and recommendation for design guidelines for healthcare usage of social media.

**Healthcare in Kuwait**

Kuwait is a small country of 18,000 sq. km. located within the Middle East region. While the population of Kuwait is around 4 million people and only 1.5 million are Kuwaiti. The small land size and small population size help the government of Kuwait to easily develop and manage its healthcare services. Despite having many private hospitals and clinics, the government is still the largest investor in the healthcare sector (Kieft et al., 2012). The healthcare system is a free service for Kuwaiti, and affordable for non-Kuwaiti. There are three categories of healthcare service providers: public (governmental), hybrid public with private (semi-governmental), and private (non-governmental) (Labor Market Information System & Kuwait Central Statistical Bureau, 2015)

The number of healthcare service providers available between 2011 and 2013 and the number of doctors serving in each category along with the number of patients. The reported numbers of doctors and patients show a growth rate in the number of patients much higher than the growth rate in the number of doctors, while the number of the healthcare facilities almost remain the same (see Table 1). Moreover, the rate of population per hospital bed is not increasing proportionally to the population growth, indicating limited coverage. The number of beds per hospital was 568, 580, and 588 for 2013, 2014, and 2015 consecutively (Kuwait Central Statistical Bureau, 2015). Thus, the regular on premises service providers might not be enough in serving the needs of the population in Kuwait. Even though, face-to-face appointments to diagnose a disease are very important to ensure the quality and accuracy of the treatments and medication prescribed to patients, some visits of patients might serve for general inquiries or follow-up and other information sharing activities which could be potentially replaced by digital communications through social media to improve information access, ensure satisfaction of the demand, reducing costs and optimizing access. The current healthcare system in Kuwait does not support any type of online consultation or any other sort of communication tool to interact with doctors directly without a face-to-face visit.

| Category | Year | Hospitals | Centers | All Doctors | Patients |
|---|---|---|---|---|---|
| **Public Sector** | 2011 | 15 | 92 | 7,349 | 2,432,773 |
| | 2012 | 15 | 94 | 7,898 | 2,495,121 |
| | 2013 | 16 | 94 | 8,691 | 3,233,456 |
| **Hybrid** | 2011 | 3 | Not available | 198 | 483,680 |
| | 2012 | 3 | Not available | 201 | 475,415 |
| | 2013 | 3 | Not available | 212 | 491,897 |
| **Private** | 2011 | 12 | Not available | 885 | 2,140,661 |
| | 2012 | 12 | Not available | 909 | 2,209,973 |
| | 2013 | 12 | Not available | 1,001 | 2,322,853 |

**Table 1:** Healthcare Service Providers in Kuwait Including Public (Governmental), Hybrid (Semi-governmental), and Private (Non-governmental) (Labor Market Information System & Kuwait Central Statistical Bureau, 2015)

The number of communicable diseases in Kuwait increased by 0.5% from 2013 to 2015 (Kuwait Central Statistical Bureau, 2015). This incidence could be reduced by educating the public to follow healthcare practices in their daily lives. Social media can improve the communication between the healthcare providers and the citizens in Kuwait by reaching out to the public in a convenient way to educate them and provide a communication tool to keep them in touch with medical practitioners. The Ministry of Health (MoH) in Kuwait adopted social media to keep the ministry updated with the latest communication technologies and to reach out to the largest possible number of citizens. MoH has a specific electronic health campaign (ehealth) that is responsible for keeping the public updated with the latest information on all online activities for the ministry in a unified view to ensure consistency among the channels. The main goal of the ehealth campaign is to create a Kuwaiti society with a high level of health awareness, ensuring that they practice healthy lifestyles independently and continuously. The communication channels used by the MoH include: an email address (info@q8-ehealth.com), a website (http://q8-ehealth.com/), a Facebook account (المكتب الإعلامي بوزارة الصحة), an Instagram account (@moh_media_office), a Twitter account (@arikuwait), and a YouTube channel ( المكتب الاعلامي لوزارة

الصحة) which are administered and managed by a group of physicians assigned by the ministry. The ehealth campaign has an office at the main building of the MoH, and a 24 hours' phone number to ensure reachability, accessibility, and reliability of their services and quick responses to customers' inquiries. The campaign provides videos, messages, and photos containing advices, medical news, other health campaigns announcements, and interviews with physicians and specialists about the latest health issues in Kuwait.

In addition to the MoH, Kuwait Medical Association (KMA) has been established in 1963, and currently having 12,000 physicians from public and private sectors. The KMA is adopting multiple social media channels. The KMA has a website (http://kma.org.kw), an email (kma@kma.org.kw), a Twitter account (@KMA_1963), a YouTube channel (kma 1963), and an Instagram account (kma_1963).

The transformation of healthcare services toward social media is increasing in Kuwait on the organizational level, including clinics, medical centers, healthcare organizations, hospitals, and on the individual professional level, including physicians, nutritionists, pharmacists etc., and mostly by younger clinicians for personal and information purposes (Von Muhlen & Ohno-Machado, 2012).

## Healthcare on Social Media

Previous studies have shown that social media have potential for broad effects on the healthcare service, transforming the behaviors of patients, medical professionals, and healthcare organizations (Taiminen, 2016; Smailhodzic et al., 2016; McCaughey et al., 2014). This shift creates new opportunities and challenges to healthcare. To the healthcare system to succeed, managing the relationship and communication between patients and medical practitioners is important, as it ensures the direct transference of information in the right time and using the right method with minimal risks and costs.

**Patients' Use of Social Media**

Patients' use of social media have been the focus for most of the recent social media studies covering healthcare. Social media offers many benefits for patients supporting the effectiveness of their treatments (Taiminen, 2016), as well as many challenges that might threaten their health if not properly investigated (Lenhard et al., 2010; Mackey et al., 2013). Patients' use of social media is classified into six categories: emotional, information, esteem, network support, social comparison, and emotional expression (Smailhodzic et al., 2016).

According to Smailhodzic et al. (2016), patients use social media to complement healthcare professional services, which helps in fulfilling their needs that cannot be met by healthcare professionals such as peer support and emotional support. Physicians often do not have enough time to discuss patients' needs in detail, while other patients feel more comfortable to discuss related matters in depth on blogs and social networking sites in a more effective, informative and educative way.

Emotional well-being is very crucial for patients in treatment. Patients use social media channels to seek emotional support, engaging in discussion with others in similar situation on online communities formed around a shared interest or goal. Social media provides emotional support for patients continuously, which play a vital role in improving their health along all stages of a treatment. Active participants in online healthcare communities perceive their ability to live healthier than did those who had not participated in such healthcare communities (Taiminen, 2016). Online active patients on Facebook or on a forum and Facebook experienced significantly more emotional support than those who are active only in a forum or did not actively participate in any social media channels (Taiminen, 2016).

Despite the benefits of social media on a patient's life, the use of social media by patients can also threaten their lives when misused. Studies show that Facebook causes Nonmedical Use of Prescription Medicines (NUPM) among youths (Lenhard et al., 2010; Mackey et al., 2013). Concerns regarding the quality of information and the reliability of sources on social media are other issues affecting patient's life (Osanlou et al. 2015).

**Medical Practitioners' Use of Social Media**

Medical practitioners can satisfy their professional needs by using social media. Social media offers many features that can be used for professional development and advancement, as well as for health education, promotion and clinical marketing, and communication activities (Bernhardt et al., 2014; Brokowski, 2009; Opel, 2016; Ventola, 2014). In addition, medical professionals can use social media to engage in discussion regarding medical practices and affect policy makers to boost the outcome of the medical services (Fogelson et al.,2013; Ventola, 2014). Medical students showed high personal use of Facebook, ranging from 64% to 96%, and professionals showed lower personal use of Facebook, ranging from 12.8% to 46.7% (Von Muhlen et al., 2012). In another study from 2009, 28% of pharmacists reported using Wikipedia for drug information, and out of them only 28% were aware that it is a crowdsource platform and anyone can edit its contents (Brokowski, 2009).

Concerns about clinicians publicizing unprofessional content or breaching patient confidentiality (Hader & Brown, 2010; Chretien et al., 2011) are the most threatening concerns regarding the professional use of social media within healthcare domain (Von Muhlen et al., 2012). The use of social media by healthcare workers can present risks under the Health Insurance Portability and Accountability Act (HIPAA) and the Health Information Technology for Economic and Clinical Health (HITECH) act (Ventola, 2014). All healthcare providers at the United States are forced to comply with HIPAA and HITECH to ensure data privacy and security provisions for safeguarding medical information and patients' confidentiality. A study conducted on Twitter data using 5156 tweets from 260 self-identified physicians, found that 4% were potentially unprofessional, including 38 potential patient privacy violations (Chretien et al., 2011).

Table 2 summarizes the benefits and challenges of social media use by medical practitioners, and patients according to an analysis of previous studies. We use the same benefits defined in Table 2 and others to guide our analysis in understanding how medical practitioners and patients are using the social media for medical purposes in Kuwait.

| Categories | Benefits | Challenges |
|---|---|---|
| **Medical Practitioners** | Professional development and advancement (Bernhardt et al., 2014; Ventola, 2014)<br>Health education (Brokowski, 2009; Bernhardt et al., 2014; Ventola, 2014)<br>Communication activities (Opel, 2016; Bernhardt et al., 2014)<br>Marketing (Opel, 2016)<br>Increase personal awareness (Ventola,2014)<br>Improve health outcome (Ventola,2014)<br>Affect health policy decisions (Fogelson et al.,2013) | Publicizing unprofessional content (Hader & Brown, 2010; Von Muhlen et al., 2012)<br>Breaching patient confidentiality (Hader & Brown, 2010; Von Muhlen et al., 2012; Chretien et al., 2011; Osanlou et al. 2015) |
| **Patient** | Emotional support (Taiminen, 2016; Smailhodzic et al., 2016)<br>Complement healthcare professional services (Smailhodzic et al., 2016; Osanlou et al. 2015)<br>Network support (Smailhodzic et al., 2016)<br>Empower individuals to manage their well-being by themselves (Osanlou et al. 2015) | Threaten patients' lives if misused (Lenhard et al., 2010; Mackey et al., 2013)<br>Lack of quality assurance (Osanlou et al. 2015) |

**Table 2:** Benefits and Challenges of Social Media Use by Patients and Medical Practitioners

## Methodology

**Design**

The International Review Board (IRB) approved this study (1131830-1/2, 1131584-1/2). Data collection consists of two online surveys. One targeted patients, and another medical practitioner. Medical practitioners have different cultures toward medical usages for social media for multiple reasons, such as: career reputation, professional policies and guidelines, which might affect their behaviors and make it differs from other users. Thus, differentiating between medical participants and others is crucial in our study. Having two samples of users including medical practitioners and patients will help for better understanding for the medical communication process on social media and give implications for designing to improve standard healthcare practices. The online surveys were administered using Google Forms.

All participants were informed about the procedures and goal of the survey before starting to fill it. The survey has four main sections: 1) assess users' familiarity with social media applications; asking about their daily frequencies of usages for multiple applications and the importance of those applications, 2) assess users' preference on multiple type of contents that are supported by most of the available social media application such as sharing pictures, videos, text, locations, and their preferences on having their contents publicly available, 3) assess the importance of multiple factors in affecting user behavior toward medical purpose, and 4) socio-demographic profile. The questions in both surveys are similar except in the third and the fourth sections, which have specific questions targeting medical professionals and patients. Questions' types include: Likert scales, open questions, closed questions, and multiple choice questions.

We sent 200 Recruitment messages by email, WhatsApp, Facebook messenger, and Instagram direct messages between October 2017 and January 2018. Sampling process includes snowball sampling (sending messaging to friends and friends of friends), and convenience sampling. Participation requirements included: reading and writing in English, and familiarity with social media platforms. We excluded only participants living in Kuwait.

**Data Analysis**

The data were analyzed using IBM Statistical Package for Social Sciences (SPSS) version 25 applying simple descriptive statistics. Answers for open questions were analyzed using coding. Answers for Likert scale questions were converted to numerical scale (always=5, often=4, sometimes=3, rarely=2, and never=1), and averages were calculated for each question. Patients' responses and medical practitioners' responses were calculated separately to measure the differences in their assessed behaviors. Responses were grouped based on participants' inputs of the most important social media platform, and results were reported to describe differences in behaviors of patients and medical practitioners on each platform.

## Results

The total number of participants was 176, including 142 patients and 34 medical practitioners. In both samples, the proportion of female was almost three times higher than the proportion of males (see Table 3). The male participants were 29% of medical practitioners and 27% of patients, while the females were 71% of medical practitioners and 72% of patients. The average age for medical practitioners was 36 (SD=7.66) with 10 (SD=6) years average number of years of experience. The average age of patient was 24 (SD=6.78). Most of medical participants were full-time employees (82%), and mostly were family physicians (27%) (see Table 4).

| Categories | Gender | frequency (%) (n=176) |
|---|---|---|
| **Medical Practitioners** | Male | 10 (29%) |
| | Female | 24 (71%) |
| | Total | 34 (19%) |
| **Patients** | Male | 40 (27%) |
| | Female | 102 (72%) |
| | Total | 142 (81%) |

**Table 3:** Gender Distribution

| Employment | Frequency (%) (n=33) |
|---|---|
| **Full-time** | 28 (82%) |
| **Student** | 3 (9%) |
| **On leave** | 1 (3%) |
| **Retired** | 1 (3%) |
| **Profession** | **Frequency (%) (n=26)** |
| **Family Physician** | 9 (26%) |
| **Dentist** | 5 (14%) |
| **Pharmacist** | 5 (14%) |
| **Medical Student** | 3 (9%) |
| **Therapist** | 1 (3%) |
| **Nutritionist** | 1 (3%) |
| **Surgeon** | 1 (3%) |
| **Dermatologist** | 1 (3%) |

**Table 4:** Medical Practitioners Demographic Profile

Participants were asked if they have any concerns regarding viewers of their account contents. Public account allows everyone with internet access to view its contents, while private account allows only authorized users to access its contents. Medical practitioners are less aware regarding their availability of their contents to the public (see Table 5). The percentage of public accounts users and not public account users were equal with 38%, while 24% of medical practitioners had no preference. Patients result shows 38% public account users, while the majority (55%) were not public account users, 5% of patients had no preference, and 2% have other preference.

| Categories | Prefer Public Account Frequency (%) (n=67) | Prefer Not Public Account Frequency (%) (n=91) | No Account Preference Frequency (%) (n=15) | Other Account Preference Frequency (%) (n=3) |
|---|---|---|---|---|
| **Medical Practitioners** | 13 (38%) | 13 (38%) | 8 (24%) | 0 (0%) |
| **Patients** | 54 (38%) | 78 (55%) | 7 (5%) | 3 (2%) |

**Table 5:** Account Preference for Viewing Contents

|  | **Medical Practitioners** | **Patients** |
|---|---|---|
| **Facebook** | Rarely | Rarely |
| **Instagram** | Often | Often |
| **LinkedIn** | Rarely | Never |
| **PatientsLikeMe** | Never | Never |
| **Snapchat** | Often | Often |
| **Twitter** | Sometimes | Often |
| **YouTube** | Often | Often |

**Table 6:** Frequency of Usage for Different Types of Social Media

Seven social media platforms were chosen to be ranked based on daily frequencies of use using a 5-point Likert scale with semantic differential (always, often, sometimes, rarely, and never). Among the seven platforms chosen, five (Facebook, Instagram, Twitter, Snapchat, and YouTube) were chosen based on a previous study on the Arab countries usage of social media which shows them as the top used ones for 2017 (Media Use in the Middle East, 2017). We added LinkedIn to get some implications on professional and career purposes use of social media. We added PatientsLikeMe to check if the participants are aware about healthcare and medical domain specific social media platforms. Facebook is rarely used by all participants. Instagram, Snapchat and YouTube are often used by all participants. Patients often use Twitter, while medical practitioners sometimes use it. Medical practitioners rarely use LinkedIn, while patients never use it. PatientsLikeMe is never used by all participants.

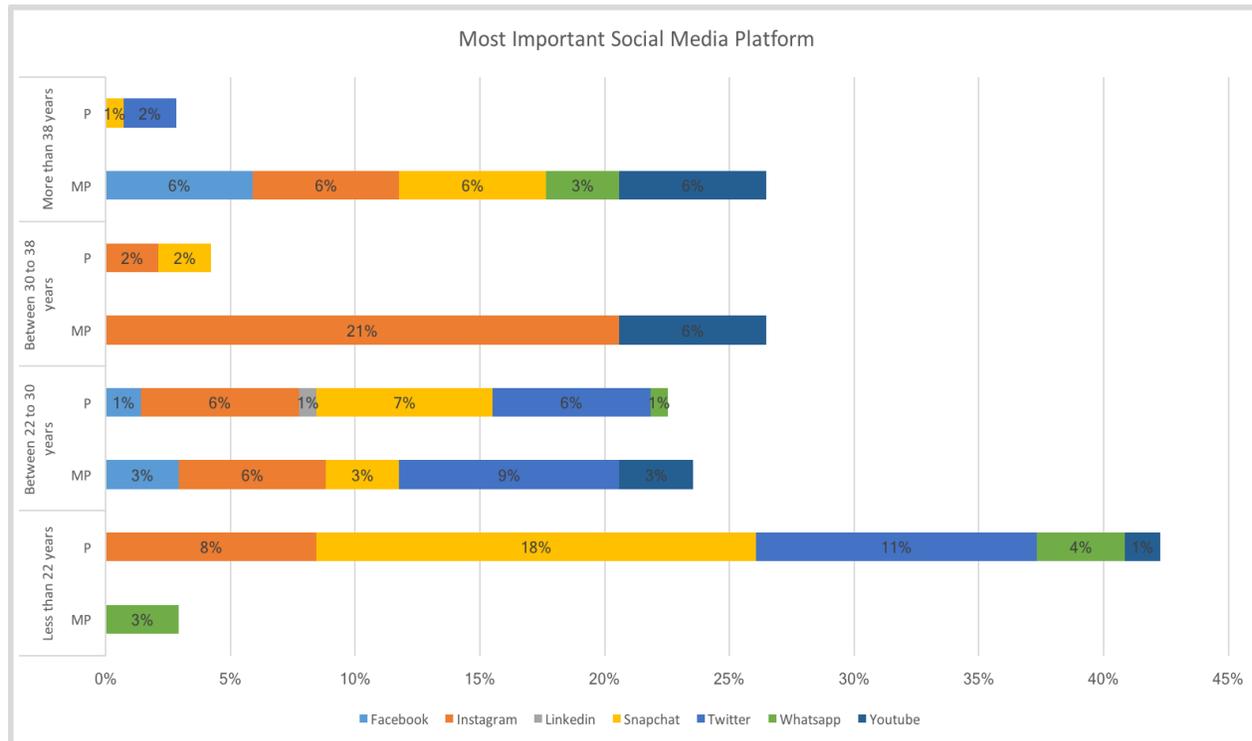

**Figure 1:** Most Important Social Media Platform Based on Age Group (MP=Medical Practitioners, P=Patients)

An open question was used to let participants choose freely their most important social media platform to guide the remaining analysis and evaluations. For young medical practitioners, whose age is less than 22, the most important social media platform is WhatsApp. Practitioners between 22 and 30 years old choose Twitter as the most important platform. Medical practitioners between 30 and 38 years rate Instagram as the most important platform. Practitioners above 38 years old give the same importance for YouTube, Facebook, Snapchat and Instagram. Young patients below 30 years old give highest importance to Snapchat. Patients whose age between 30 and 38 years, give equal importance to Instagram and Snapchat. Patients above 38 years old, rate Twitter as the most important platform (see Figure 1).

| Activities | Facebook | Instagram | LinkedIn | Snapchat | Twitter | WhatsApp | YouTube |
|---|---|---|---|---|---|---|---|
| **Comment on others posts** | Sometimes | Sometimes | Sometimes | Sometimes | Often | Sometimes | Sometimes |
| **Like others posts** | Often | Often | Rarely | Sometimes | Often | Sometimes | Sometimes |
| **Share a picture** | Often | Often | Never | Always | Sometimes | Often | Sometimes |
| **Share a text message** | Sometimes | Sometimes | Rarely | Rarely | Often | Always | Sometimes |
| **Share a video** | Sometimes | Sometimes | Never | Always | Rarely | Often | Sometimes |
| **Share contents from other sources** | Sometimes | Sometimes | Sometimes | Often | Often | Sometimes | Sometimes |
| **Share my location** | Never | Rarely | Rarely | Rarely | Rarely | Rarely | Rarely |

**Table 7:** Type of Contents with Their Frequency of Sharing on Social Media by Medical Professionals

Participants were asked to rate their daily frequencies of sharing multiple type of contents (e.g. picture, video, text) and participating in online activities (e.g. liking, commenting) on their most important platform using a 5-point Likert scale with semantic differential (always, often, sometimes, rarely, and never). Medical practitioners who use Facebook and Instagram are on the same state, they often share pictures and like others' posts, they sometimes share video, text messages, comment on others' posts, share content from other sources, but they never share their location. Practitioners who use LinkedIn are the least active ones, they never share pictures and videos, they rarely share text message, like other's posts, and share location, and they sometimes comment on others post and share contents from other sources. Practitioners who use Snapchat are always sharing pictures and videos, they often share contents from other sources, they sometimes comment on other's posts and likes other's posts, and they rarely share text message and location. Practitioners who use Twitter are the most active. They often share text message, comments on other's posts, likes other's posts, and share contents from other sources, they sometimes share pictures, and they rarely share videos and their location. Practitioners who use

WhatsApp always share text message, they often share picture and video, and they rarely share location. While practitioners who use YouTube are on average on almost all activities. They sometimes share picture, video, text message, comments on other's posts, likes other's posts, and share contents from other sources, while they rarely share location (see Table 7).

| Activities | Facebook | Instagram | LinkedIn | Snapchat | Twitter | WhatsApp | YouTube |
|---|---|---|---|---|---|---|---|
| **Comment on others posts** | Often | Sometimes | Sometimes | Sometimes | Sometimes | Sometimes | Sometimes |
| **Like others posts** | Often | Sometimes | Sometimes | Sometimes | Sometimes | Often | Sometimes |
| **Share a picture** | Always | Often | Never | Often | Sometimes | Often | Often |
| **Share a text message** | Often | Often | Sometimes | Often | Often | Often | Often |
| **Share a video** | Always | Sometimes | Never | Sometimes | Sometimes | Often | Often |
| **Share contents from other sources** | Always | Sometimes | Rarely | Sometimes | Rarely | Sometimes | Often |
| **Share my location** | Often | Rarely | Never | Rarely | Rarely | Sometimes | Sometimes |

**Table 8:** Type of Contents with Their Frequency of Sharing on Social Media by Patients

Patients who use Facebook are very active on online activities, they always share pictures, videos, and contents from other sources, while they often share text message, comments on others posts, like others' posts and share location. Patients who use Instagram and Snapchat are on the same online activities level, they often share picture and text message, they sometimes share videos, comment on other's post, like other's posts, and share contents from other sources, and they rarely share location. Patients who use LinkedIn sometimes share text message, comment on other's posts, and like others post, they rarely share contents from other sources, and they never share picture, video and location. Patients on Twitter often share text message, they sometimes share pictures, videos, comments on others post, and they rarely share contents from other sources and location. Patients on WhatsApp and YouTube are similar in their activities, they often share pictures, videos, text messages, and they sometimes share comments on other's

18posts and location. While WhatsApp users often like others' posts, YouTube' users are sometime liking others' posts. YouTube users often share contents from other sources, while WhatsApp users sometime sharing contents from other sources (see Table 8).

Benefits of social media for patients and medical practitioners defined on Table 2 were examined further in our study. Table 9 shows the questions used to assess benefits of social media for medical practitioners in the survey. We added questions to examine privacy concerns regarding the terms of use of platforms currently in use. We added another question to assess the contributions of medical practitioners as a source of medical contents.

| Questions for Medical Practitioners | Categories |
|---|---|
| I use my social media accounts to connect with my clients. | Communication activities |
| I use my social media accounts to learn about healthcare and medical issues. | Health education |
| I use my social media accounts to connect with medical news outlets. | Health education |
| I use my social media accounts to search for information on medicines, diseases, and treatments. | Health education |
| I use my social media accounts to announce the services offered in my clinic. | Marketing |
| I use my social media accounts to connect with other people within the medical field. | Network support |
| I read the terms of use for social media application that I am using now. | Privacy concerns |
| I use my social media accounts to learn about the services on other clinics. | Professional development |
| I post medical information on my social media accounts. | Source of medical contents |

**Table 9:** Questions Used to Measure the Behavior of Medical Practitioners on Social Media

| Categories | Facebook | Instagram | LinkedIn | Snapchat | Twitter | WhatsApp | YouTube |
|---|---|---|---|---|---|---|---|
| Communication Activities | Rarely | Rarely | Rarely | Sometimes | Rarely | Rarely | Rarely |
| Health Education Purposes | Sometimes | Sometimes | Sometimes | Sometimes | Sometimes | Sometimes | Sometimes |
| Marketing | Often | Rarely | Sometimes | Sometimes | Rarely | Rarely | Rarely |
| Network Support | Sometimes | Sometimes | Sometimes | Rarely | Often | Sometimes | Sometimes |
| Privacy Concerns | Sometimes | Rarely | Sometimes | Rarely | Rarely | Rarely | Sometimes |
| Professional Development | Sometimes | Sometimes | Sometimes | Sometimes | Rarely | Rarely | Sometimes |
| Source of Medical Contents | Sometimes | Rarely | Sometimes | Sometimes | Rarely | Rarely | Rarely |

**Table 10:** Medical Practitioners Behavior on Social Media Toward Healthcare Contents Based on Social Media Platforms

The results on Table 10 report the behavior of medical practitioners on social media per social media platform. Medical practitioners who use Facebook and LinkedIn are quite similar in some attributes, they both sometimes become sources of medical information, health education purposes, network support, professional development, and have privacy concerns, and they rarely use the platforms for communication activities. Facebook users are often using it for marketing, while LinkedIn users sometimes use it for marketing. Instagram, Twitter and WhatsApp users are sharing some similarities, they rarely become sources of medical contents, communication activities and marketing, and they sometimes used social media for health education. Instagram, YouTube, and WhatsApp are sometimes used for network support, while Twitter is often used for network support. Twitter and WhatsApp are rarely used for increasing professional development, and their users are not aware of privacy concerns. Users of Instagram and YouTube sometimes use the platform for professional development. Instagram users are rarely aware of privacy issues, while YouTube users are sometimes aware. Medical practitioners who use Snapchat are sometimes using it to serve almost all purposes except for network support, and they rarely have privacy concerns.

| Questions for Patients | Categories |
|---|---|
| I use my social media account to learn about the services offered by clinics before going to anyone. | Complement healthcare professional services |
| When I got sick, I tried to search social media for others in similar medical situation. | Emotional support |
| I use my social media account to learn about healthcare and medical issues | Health education |
| I use my social media account to connect with medical news outlets | Health education |
| I use my social media account to search for medicines, diseases, and treatments. | Health education |
| I use my social media account to connect with my friends | Network support |
| I use my social media account to connect with other people in similar profession of mine. | Network support |
| I use my social media account to connect with family. | Network support |
| I read the terms of use for the social media application that I am using now | Privacy Concerns |
| Looking for solutions to my medical problem in the social media can replace my visit to the doctor. | Replace healthcare professional services |
| I post medical information on my social media accounts. | Source of medical contents |

**Table 11:** Questions Used to Measure the Behavior of Patients on Social Media

| Categories | Facebook | Instagram | LinkedIn | Snapchat | Twitter | WhatsApp | YouTube |
|---|---|---|---|---|---|---|---|
| Complement Healthcare Professional Support | Often | Sometimes | Never | Sometimes | Sometimes | Sometimes | Sometimes |
| Emotional Support | Often | Sometimes | Never | Sometimes | Sometimes | Rarely | Often |
| Health Education Purposes | Often | Sometimes | Rarely | Sometimes | Sometimes | Often | Sometimes |
| Network Support | Often | Often | Sometimes | Often | Sometimes | Often | Often |
| Privacy Concerns | Sometimes | Sometimes | Rarely | Rarely | Rarely | Sometimes | Often |
| Replace Healthcare Professional Services | Often | Sometimes | Never | Sometimes | Rarely | Never | Sometimes |
| Source of Medical Contents | Often | Rarely | Never | Rarely | Rarely | Rarely | Sometimes |

**Table 14:** Patient Behavior on Social Media Toward Healthcare Contents Based on Social Media Platforms

Patients who use Instagram share some similarities with those who use Snapchat, they sometimes use the platform for health education purposes, complementing healthcare professional support, emotional support, and replacing healthcare professional services, they rarely become sources for medical contents, and they often use the platform for network support. Instagram users sometimes have privacy concerns, while Snapchat users rarely have. Facebook users often use the platform to receive all benefits except that they sometimes have privacy concerns. Patients who use LinkedIn are inactive regarding medical purposes, they never use it to receive most of the medical purposes, they rarely use it for health education purposes, and they sometimes use it for network support. Twitter users sometimes use it to receive all medical purposes, except they rarely become a source of medical contents, replace healthcare professional services, and have privacy concerns. Patients often use WhatsApp for healthcare education and network support, they sometime use it to complement healthcare professional support, and have privacy concerns. WhatsApp users rarely contribute as a source of medical contents and seek emotional support, and they never replace healthcare professional service with WhatsApp. YouTube users are sometimes contributing as a source of medical contents, use it for health education purpose, complement healthcare professional support, and replace healthcare professional support (raising patients' safety concerns). Patients on YouTube often seek network support, emotional support, and have privacy concerns.

## Discussion

**Key Findings**

The results of our study show slight differences in the behaviors and awareness within medical contexts for social media use by patients and medical practitioners. The profile characteristics of users of social media for healthcare shows slightly higher medical purposes use by patients than medical practitioners, in general among multiple platforms. However, both patients and medical practitioners are not using the current available social media platforms for medical purposes with full potential benefits, they sometimes use social media to fulfil medical purposes. Even though, social media is very important for them and they are using it daily, but most of their usages are not for medical purposes. This result is compatible with previous studies, which reported ordinary usage of social media by medical practitioners (McGowan et al., 2012)

Specifically, patients reported slightly more awareness than practitioners in health education, health-related network support, and communication activities. While, medical practitioners are slightly more aware than patients to use social media as a source of medical information, for clinician marketing and for professional development. Even though all participants reported rare usage of Facebook, patients between 22 and 30 years and medical practitioners above the age of 38 years, who are using Facebook, seem to be the most active users in online activities for medical purposes. Previous studies show that young medical professionals are more likely to use social media for medical purposes (Von Muhlen & Ohno-Machado, 2012), but our results show that middle age and old professionals are more likely to use social media for medical purposes. Rating Facebook' users as the most active users is compatible with findings from previous studies for medical practitioners (Von Muhlen & Ohno-Machado, 2012) and for governmental employees in Kuwait (Chaudhry, 2013).

Medical Practitioners are slightly less aware than patients regarding privacy settings regarding the platform's term of use (rarely for medical practitioners and sometimes for patients) and availability of their contents for the public (public account and not public account users are both 38% for medical practitioners, while 38% are public account users and 55% are not public account users for patients). Our findings regarding the availability of account contents to the public differ from previous study which reported tribal behavior by clinicians in sharing their domain knowledge, clinicians from one organization do not share their knowledge with others belonging to other organizations (Rolls et al., 2016).

Overall, patients and medical practitioners show similar engagement rates on multiple online activities (e.g., share pictures, videos, text message) among platforms, except for sharing text message which is more practiced by patients. Sharing text messages is the mostly practiced activity by patients, while medical practitioners mostly share contents from other sources. Location shows significant low sharing rates among all participants. Patients using Facebook are the most active patients, while medical practitioners using Snapchat are the most active medical practitioners in term of engaging in online activities and in comparison with other platforms. Our findings match with the findings from a previous study targeted at medical professional usage of social media, which showed low posting behaviors but more frequent reading and accessing behaviors (Rolls et al., 2016).

**Implications for Designing Social Media Platform to Serve Healthcare Purposes**

The findings highlighted the need to increase the awareness of all social media users toward healthcare and medical purposes. Healthcare organizations should establish online campaigns to increase the sources and medical contents on platforms, and to encourage other users to circulate the medical contents, particularly medical practitioners. Platform design should focus on factors serving the main needs for medical practitioners by adding their career identity to their profile to foster professional development among their international peers. Adding features for medication and medical repository or dictionary can serve patients and professionals. Connecting patients by creating online communities within platforms can serve their needs for emotional and network support. Moreover, having a social privacy setting that let users autonomously set who view their account contents based on each post rather than the entire account can increase their engagements by solving their privacy concerns especially for patients.

## Limitations

Results may be generalizable with caution. The gender distribution for patients and medical practitioners is not proportional to the gender distribution in Kuwait. On 2017, the number of females in Kuwait was 1,663,866, and males was 2,773,724, while on 2016, the percentage of female employees in the governmental sector was 52.4% and male was 47.5% (The Public Authority for Civil Information, 2017). Thus, our findings might be appropriate to generalize on employees in the governmental sector.

24## Conclusions

This study has demonstrated the need to increase healthcare awareness within social media users in Kuwait. Patients and medical practitioners are not using the current social media channels for healthcare purposes. The current healthcare system in Kuwait has problems, the number of patients is increasing while the number of healthcare service providers is unchanged. This problem could be solved or reduced if the current social media channels were used to serve societal medical needs. The design of social media platform can help its users to engage in activities serving medical purposes if it considers users' needs in platform design.

## Future work

Our study investigates the current behavior of medical practitioners and patients in social media for medical purposes in Kuwait. We did not consider the factors behind the current uses. It would be helpful to examine the factors behind the general reduction in healthcare and medical awareness in Kuwait. Knowing the causes can help in better planning to transform the use toward more positive healthcare purposes.

## References

AlAjlan, M. (2011). Students' attitudes towards communicating with instructors via internet: a case study in the College of Engineering & Petroleum: Kuwait University. *In Proceedings of the Second Kuwait Conference on e-Services and e-Systems (KCESS '11)*, 6, 1:4. DOI: https://doi.org/10.1145/2107556.2107562

AlHuwail, D. & Barnes, R. (2011). Diabetes care in the age of informatics: Kuwait-Scotland health innovation network. *In Proceedings of the Second Kuwait Conference on e-Services and e-Systems (KCESS '11)*, 16, 1:7. DOI: https://doi.org/10.1145/2107556.2107572